\documentclass{eptcs}

\usepackage{amsthm}

\newtheorem{theorem}{Theorem}
\newtheorem{lemma}{Lemma}
\newtheorem{proposition}{Proposition}
\newtheorem{corollary}{Corollary}
\theoremstyle{definition}
\newtheorem{example}{Example}

\usepackage{stmaryrd} 

\usepackage{amsmath, amssymb}
\usepackage{graphicx}
\usepackage{bussproofs}
\EnableBpAbbreviations
\usepackage{upgreek}

\usepackage{natbib}
\bibpunct();A{},

\usepackage{textcomp}

\hypersetup{
pdfauthor   = {Daniel Seidel <ds@iai.uni-bonn.de> and Janis Voigtl\"ander <jv@iai.uni-bonn.de>},
pdftitle    = {Improvements for Free},
pdfkeywords = {functional programming, polymorphism, relational parametricity, logical relations, free theorems, efficiency}}

\providecommand{\event}{QAPL 2011}

\title{Improvements for Free}
\author{Daniel Seidel\thanks{This author was supported by the DFG under grant VO 1512/1-1.} \qquad \qquad \qquad \qquad Janis Voigtl\"ander
\institute{Rheinische Friedrich-Wilhelms-Universit\"at Bonn,
Institut f\"ur Informatik\\
R\"omerstra{\ss}e 164, 53117 Bonn, Germany}
\email{\{ds,jv\}@informatik.uni-bonn.de}
}
\def\titlerunning{Improvements for Free}
\def\authorrunning{D.\ Seidel \& J.\ Voigtl\"ander}

\newcommand{\pbr}{$ $}
\newcommand{\ie}{i.e.,\ }
\newcommand{\cif}{\text{if }}

\newcommand{\with}{\text{~~with }}
\newcommand{\eqa}{\ensuremath{\Leftrightarrow}}
\newcommand{\impl}{\ensuremath{\Leftarrow}}
\newcommand{\fspace}[2]{{#2}^{#1}}

\newcommand{\sem}[1]{\ensuremath{\llbracket #1 \rrbracket}} 
\newcommand{\ctsem}[1]{\ensuremath{\llbracket #1 \rrbracket'}} 
\newcommand{\csem}[1]{\ensuremath{\llbracket #1 \rrbracket^\text{\rm{\textcent}}}} 
\newcommand{\getVal}[1]{\ensuremath{\mathit{val}(#1)}}
\newcommand{\getCost}[1]{\ensuremath{\mathit{cost}(#1)}}
\newcommand{\Cost}[1]{\ensuremath{\mathcal{C}(#1)}}
\newcommand{\cOpApp}{\ensuremath{\mathrel{\text{\rm{\textcent}}}}} 
\newcommand{\cOpCons}{\ensuremath{\mathrel{:^\text{\rm{\textcent}}}}} 
\newcommand{\cOpPair}[2]{\ensuremath{(#1, #2)^\text{\rm{\textcent}}}} 
\newcommand{\clist}[1]{\ensuremath{[#1]^\text{\rm{\textcent}}}} 

\newcommand{\appCost}{\ensuremath{\mathit{appCost}}}

\newcommand{\Z}{\ensuremath{\mathbb{Z}}}
\newcommand{\N}{\ensuremath{\mathbb{N}}}
\newcommand{\datasyntax}[1]{\ensuremath{\mathsf{#1}}}
\newcommand{\textsyntax}[1]{\ensuremath{\mathbf{#1}}}
\newcommand{\var}[1]{\ensuremath{\mathit{#1}}}
\newcommand{\Int}{\datasyntax{Nat}}

\newcommand{\sv}[1]{\ensuremath{\mathbf{#1}}} 
\newcommand{\svv}[1]{\ensuremath{\mathbf{#1}}} 
\newcommand{\svc}[1]{\ensuremath{#1}} 

\newcommand{\id}{\ensuremath{\var{id}}}

\newcommand{\nil}{\ensuremath{[\,]}} 
\newcommand{\vnil}{\ensuremath{\sv{[\,]}}} 
\newcommand{\cons}{\ensuremath{:}}

\newcommand{\plFold}{\ensuremath{\mathbf{lfold}}}
\newcommand{\lFold}[1]{\ensuremath{\plFold(#1)}}
\newcommand{\iFold}[1]{\ensuremath{\mathbf{ifold}(#1)}}
\newcommand{\tPair}[2]{\ensuremath{(#1,#2)}}

\newcommand{\mapPair}{\ensuremath{\var{mapPair}}}
\newcommand{\vmapPair}{\ensuremath{\sv{mapPair}}}
\newcommand{\mapList}{\ensuremath{\var{mapList}}}
\newcommand{\vmapList}{\ensuremath{\sv{mapList}}}

\newcommand{\cta}{\ensuremath{::}}
\newcommand{\slogrel}[2]{\ensuremath{\Delta_{#1, #2}}}
\newcommand{\logrel}[2]{\ensuremath{\plogrel_{#1, #2}}}
\newcommand{\plogrel}{\ensuremath{\Delta'}}
\newcommand{\clogrel}[2]{\ensuremath{\Delta^\text{\rm{\textcent}}_{#1, #2}}}

\newcommand{\tsone}{\sigma_1}
\newcommand{\tstwo}{\sigma_2}
\newcommand{\plift}[2]{\ensuremath{\mathit{lift}_{( , )}(#1, #2)}}
\newcommand{\llift}[1]{\ensuremath{\mathit{lift}_{\nil}(#1)}}
\newcommand{\cplift}[2]{\ensuremath{\mathit{lift}^\text{\rm{\textcent}}_{( , )}(#1, #2)}}
\newcommand{\cllift}[1]{\ensuremath{\mathit{lift}^\text{\rm{\textcent}}_{\nil}(#1)}}

\newcommand{\gcase}[5]{\textsyntax{case }~#1~\textsyntax{ of }~\{#2\to#3\,; #4 \to #5\}}
\newcommand{\icase}[4]{\gcase{#1}{0}{#2}{#3}{#4}}
\newcommand{\lcase}[5]{\gcase{#1}{\nil}{#2}{#3 : #4}{#5}}
\newcommand{\pcase}[4]{\textsyntax{case }~#1~\textsyntax{ of }~\{(#2, #3) \to #4 \} }
\newcommand{\ts}{\sigma}
\newcommand{\ets}{\ensuremath{\eset}}
\newcommand{\eset}{\ensuremath{\emptyset}}

\newcommand{\cApp}{\svc{1}}
\newcommand{\addCost}[2]{\ensuremath{#1 \mathop{\vartriangleright} #2}}

\newcommand{\rel}{\ensuremath{\mathit{Rel}}}
\newcommand{\rR}{\ensuremath{R}}
\newcommand{\xs}{\var{xs}}

\begin{document}
\maketitle

\begin{abstract}
  ``Theorems for Free!'' \citep{wad89} is a slogan for a technique that allows to derive statements about functions just from their types.
  So far, the statements considered have always had a purely extensional flavor: statements relating the value semantics of program expressions, but not statements relating their runtime (or other) cost.
  Here we study an extension of the technique that allows precisely statements of the latter flavor, by deriving quantitative theorems for free.
  After developing the theory, we walk through a number of example derivations.
  Probably none of the statements derived in those simple examples will be particularly surprising to most readers, but what \emph{is} maybe surprising, and at the very least novel, is that there is a general technique for obtaining such results on a quantitative level in a principled way.
  Moreover, there is good potential to bring that technique to bear on more complex examples as well.
  We turn our attention to short-cut fusion \citep{glp93} in particular.
\end{abstract}

%

\section{Introduction}\label{sec:intro}

Based on the concept of relational parametricity~\citep{rey83}, \citet{wad89} established so-called ``free theorems'', a method for obtaining proofs of program properties from parametrically polymorphic types in purely functional languages.
For example, it can thus be shown that every function $f :: [\alpha]\to[\alpha]$, with $\alpha$ a type variable, satisfies
\begin{equation}
  \label{eq:rev}
  f~(\mapList~g~\xs)=\mapList~g~(f~\xs)
\end{equation}
for every choice of $g::\tau_1\to\tau_2$ and $\xs::[\tau_1]$, with $\tau_1$ and $\tau_2$ concrete types, where:
\[
\begin{array}{@{}l@{}}
  \mapList \cta (\alpha \to \beta) \to [\alpha] \to [\beta]\\
  \begin{array}{@{}l@{\;=\;}l@{}}
    \mapList~g~[\,]    & [\,]\\
    \mapList~g~(x:\xs) & (g~x) : (\mapList~g~\var{xs})
  \end{array}
\end{array}
\]
Statements of that flavor have been used for program transformation~\citep{glp93, sve02, voi09a}, but also for other interesting results~\citep{voi08,bjc10}.

So far, free theorems have been considered a qualitative tool only.
That is, statements like~(\ref{eq:rev}) have been established as extensional equivalences or semantic approximations in a definedness order, and in fact a lot of research has gone into what definedness and/or strictness conditions are needed on the involved functions in various language settings and into extending the approach to richer type systems~\citep{laupat96,jv04,sv09,voi09b,csv10a,bjp10}.
It is natural, though, to ask about the quantitative content of free theorems in terms of program efficiency.
In a statement like~(\ref{eq:rev}), what is the relative performance of the left- and right-hand sides?
If we can answer such questions formally, this will clearly be of particular interest for the mentioned program transformation applications, where statements about efficiency have so far only been made informally or empirically.

In this paper, we lay the ground for formal such investigations.
The challenge, of course, as for standard free theorems, is to work independently of concrete function definitions, just as~(\ref{eq:rev}) depends on \emph{only the type} of $f$.  
To this end, we revise the theory of relational parametricity, essentially marrying it with the classical idea of externalizing the intensional property ``computation time'' by making it part of the observable program output, and thus accessible to semantic analysis~\citep{wad88,bjehol89,ros89,san95}.
Our vision is to eventually integrate our results into a tool like \url{http://www-ps.iai.uni-bonn.de/cgi-bin/free-theorems-webui.cgi} to enable automatic generation of quantitative free theorems for realistic languages.

To start simple, let us consider some examples.
We begin with $f :: \alpha\to\Int$.
The standard free theorem derived from that type is that for every $g::\tau_1\to\tau_2$ and $x::\tau_1$,
\begin{equation}
  \label{eq:const}
  f~(g~x) = f~x
\end{equation}
In fact, absent nontermination, it is even possible to conclude that $f$ is a constant function, i.e., for some $n::\Int$, $f$ is semantically equivalent to $(\lambda x \to n)$.
If we take program runtime into account, then there is another degree of freedom, in addition to picking the natural number $n$.
Namely, two functions of type $\alpha\to\Int$ can then differ in how long they take before providing their output, because clearly a function that no matter what the input is immediately returns $42$ is to be considered different from one that does the same after 7\textonehalf\ million years.
Even so, since the same $f$ occurs on the left- and right-hand sides of~(\ref{eq:const}), we can intuitively argue that the right-hand side will never be less efficient than the left-hand side (while it may be more efficient in that it avoids an application of $g$).
On the extensional semantics level, such invariance, namely that different $f$ may use different $n$ in $(\lambda x \to n)$, but the different instantiations of a single polymorphic $f$ at the types $\tau_2$ and $\tau_1$ on the left- and right-hand sides of~(\ref{eq:const}) may not, is exactly what relational parametricity provides.
Our task is to formally transfer this argument to the mentioned second degree of freedom, concerning program runtime.

As soon as we do consider runtime, we also have to talk about evaluation order.
For the example~(\ref{eq:const}), we can make more precise statements if we know whether function application is call-by-value or call-by-name/need.
In the former, strict case, the right-hand side of~(\ref{eq:const}) is actually more efficient than the left-hand side, because the very real cost of applying $g$ is saved.
In nonstrict languages, in contrast, the left- and right-hand sides of~(\ref{eq:const}) are to be considered equally efficient since from the type of $f$ we claimed that the function never looks at its argument (extensionally $f = (\lambda x \to n)$ for some arbitrary but fixed $n$), so the potentially costly inner application $(g~x)$ on the left-hand side is never actually evaluated.
Such issues, and the required reasoning, become more interesting as the types considered get more complicated.
For example, for the type $f :: \alpha\to\alpha\to\alpha$ and the associated free theorem
\begin{equation}
  \label{eq:proj}
  f~(g~x)~(g~y) = g~(f~x~y)
\end{equation}
the situation is the same as for~(\ref{eq:const}), i.e., the right-hand side is more efficient in a call-by-value language, while no difference is observable with call-by-name/need.
But for the type $f :: \alpha\to(\alpha,\alpha)$ and free theorem
\begin{equation}
  \label{eq:pair}
  f~(g~x) = \mapPair~(g,g)~(f~x)
\end{equation}
where
\[
\begin{array}{@{}l@{}}
  \mapPair \cta (\alpha \to \gamma, \beta \to \delta) \to (\alpha, \beta) \to (\gamma, \delta)\\
  \mapPair~(f_1,f_2)~(x_1,x_2) = (f_1~x_1, f_2~x_2)\\
\end{array}
\]
the situation is rather different: under call-by-value and call-by-need the left-hand side is more efficient, while under call-by-name the left-hand side is for sure not less efficient than the right-hand side, but whether it is actually more efficient depends on what runtime cost we associate with $\mapPair$.\footnote{In principle, one could replace $\mapPair~(g,g)~(f~x)$ by $\mathbf{let}~(y_1,y_2)=f~x~\mathbf{in}~(g~y_1,g~y_2)$ and consider $\mathbf{let}$-binding to be cost-neutral, in which case $f~(g~x)$ and the given replacement would be equally efficient under call-by-name. For call-by-value and call-by-need such replacement has no real impact, since for them a whole application of $g$ is saved on the left in any case.}
In summary, the relationships between the runtimes of the various left- and right-hand sides claimed above are as follows:
\begin{center}
  \begin{tabular}{@{\,}c|c|c|c@{\,}}
    & $f :: \alpha\to\Int$ & $f :: \alpha\to\alpha\to\alpha$ & $f :: \alpha\to(\alpha,\alpha)$ \\
    \cline{2-4} & & \\[-2.5ex]
    & $f~(g~x) = f~x$ & $f~(g~x)~(g~y) = g~(f~x~y)$ & $f~(g~x) = \mapPair~(g,g)~(f~x)$\\
    \hline & & \\[-2.5ex]
    \begin{tabular}{@{}c@{}}call-by-value\end{tabular}
    &  lhs $>$ rhs &  lhs $>$ rhs &  lhs $<$ rhs \\
    \hline & & \\[-2.5ex]
    \begin{tabular}{@{}c@{}}call-by-name\end{tabular}
    &  lhs $=$ rhs &  lhs $=$ rhs &  lhs $\leq$ rhs\\
    \hline & & \\[-2.5ex]
    \begin{tabular}{@{}c@{}}call-by-need\end{tabular}
    &  lhs $=$ rhs &  lhs $=$ rhs &  lhs $<$ rhs
  \end{tabular}
  \vspace{1ex}
\end{center}

In this paper we concentrate on call-by-value.
From the above, one could jump to the conclusion that then the answer to the question which of the two sides of a free theorem is more efficient depends only on the numbers of syntactic occurrences of $g$.
However, this simplistic view breaks down if one considers types that allow more diverse behavior, like $f :: \alpha\to\alpha\to(\alpha,\alpha)$ or indeed example~(\ref{eq:rev}).
Also, even for the cases considered above, one should not be deceived by the apparent obviousness of the analysis.
For example, that any function $f :: \alpha\to\alpha\to\alpha$ is, by its type alone, not only forced to extensionally be one of the two possible (curried) projections (a fact that can be proved using standard free theorems), but also prevented from causing different costs in different concrete invocations is a nontrivial property that requires proof.
To emphasize this point, consider a function $f :: \Int\to\Int\to\Int$.
Even if we knew that extensionally this function is equivalent to either $(\lambda x~y \to x)$ or $(\lambda x~y \to y)$, or even if we knew to which of the two, there would be absolutely no way to conclude which if any of $f~(g~x)~(g~y)$ and $g~(f~x~y)$ is more efficient for general $g::\Int\to\Int$ and $x,y::\Int$.\footnote{For example, $f$ could be a function that first counts down its first argument to zero, before finally returning its second argument. Then, by choosing $g$ and $x$ appropriately, one could make either of $f~(g~x)~(g~y)$ and $g~(f~x~y)$ arbitrarily more costly while not affecting the other one at all.}
It is only the polymorphism in $f :: \alpha\to\alpha\to\alpha$ that allows such analysis, and what we seek here is the appropriate formal theory as opposed to just the suggestive examples given above.

While the above table may suggest that we are going to prove only comparative statements, actually we will be able to make more precise quantitative statements about the relative costs of left- and right-hand sides of free theorems.
For example, for $f :: \alpha\to\alpha\to\alpha$, in the call-by-value setting, we will not only deduce that the left-hand side $f~(g~x)~(g~y)$ takes more time than the right-hand side $g~(f~x~y)$, but will also obtain that the cost difference is exactly either the cost of applying $g$ on $x$ (without the cost of evaluating $x$ itself) or the cost of applying $g$ on $y$ (without the cost of evaluating $y$ itself).

\section{A polymorphically typed lambda-calculus}\label{sec:calculus}

For formal investigation, we use a relatively small toy language that nevertheless captures essential aspects relevant for our intended analysis.
The syntax and typing rules are given in Figures~\ref{fig:syntax} and~\ref{fig:typingrules}, respectively.
There, $\alpha$ ranges over type variables, $x,y$ over term variables, and $n$ over the naturals.
The language is explicitly typed, the notation for type annotations is ``$\cta$'', while ``$:$'' is the cons operator for lists.
The operators $\mathbf{lfold}$ (corresponding to Haskell's $\var{foldr}$) and $\mathbf{ifold}$ are used to express structural recursion on lists and naturals, respectively.
(General, potentially nonterminating, recursion is not included for simplicity.)
For example, the function $\mapList$ from the introduction is defined in our calculus as follows:
\[\mapList = \lambda \mathit{g} \cta (\alpha \to \beta). \lambda \var{ys} \cta [\alpha]. \lFold{\lambda x \cta \alpha. \lambda \var{xs} \cta [\beta]. (g~x) : \var{xs}, \nil_{\beta}, \var{ys}}\]
and satisfies $\alpha,\beta\vdash \mapList \cta (\alpha \to \beta) \to [\alpha] \to [\beta]$.
\begin{figure}
  \centering
  $\begin{array}{@{}r@{\;}c@{\;}l@{}}
    \tau & ::= & \alpha \mid \Int \mid (\tau, \tau) \mid [\tau] \mid \tau \to \tau \\
    t    & ::= & x \mid n \mid \icase{t}{t}{x}{t} \mid t + t \mid \nil_{\tau} \mid t \cons t \mid \lcase{t}{t}{x}{x}{t} \mid \\
         &     & (t, t) \mid \pcase{t}{x}{x}{t} \mid \lambda x \cta \tau. t \mid t~t \mid \lFold{t,t,t} \mid \iFold{t,t,t}
  \end{array}$
  \caption{Syntax of the calculus}
  \label{fig:syntax}
\end{figure}
\begin{figure}
  \centering
  $\Gamma,x\cta\tau\vdash x\cta\tau
  $
  \qquad 
  $\Gamma\vdash n \cta \Int
  $
  \qquad
  $\Gamma\vdash[\,]_{\tau} \cta [\tau]
  $\\[2ex]

  
  \AXC{$\Gamma\vdash t_1\cta\Int$}
  \AXC{$\Gamma\vdash t_2\cta\Int$}
  \BIC{$\Gamma\vdash (t_1+t_2)\cta\Int$}
  \DP
  \qquad\qquad
  \AXC{$\Gamma \vdash t\cta\Int$}
  \AXC{$\Gamma \vdash t_1\cta\tau$}
  \AXC{$\Gamma, x \cta \Int \vdash t_2\cta\tau$}
  \TIC{$\Gamma\vdash (\icase{t}{t_1}{x}{t_2})\cta\tau$}
  \DP\\[2ex]

  
  \AXC{$\Gamma\vdash t_1 \cta \tau$}
  \AXC{$\Gamma\vdash t_2 \cta [\tau]$}
  \BIC{$\Gamma\vdash (t_1:t_2) \cta [\tau]$}
  \DP
  \qquad\qquad
  \AXC{$\Gamma \vdash t \cta [\tau_1]$}
  \AXC{$\Gamma \vdash t_1 \cta \tau$}
  \AXC{$\Gamma, x \cta \tau_1, y \cta [\tau_1] \vdash t_2 \cta \tau$}
  \TIC{$\Gamma \vdash (\lcase{t}{t_1}{x}{y}{t_2}) \cta \tau$}
  \DP\\[2ex]

  \AXC{$\Gamma\vdash t_1\cta\tau_1$}
  \AXC{$\Gamma\vdash t_2\cta\tau_2$}
  \BIC{$\Gamma\vdash (t_1, t_2)\cta (\tau_1, \tau_2)$}
  \DP
  \qquad\qquad
  \AXC{$\Gamma \vdash t \cta (\tau_1, \tau_2)$}
  \AXC{$\Gamma, x \cta \tau_1, y \cta \tau_2 \vdash t_1 \cta \tau$}
  \BIC{$\Gamma \vdash (\pcase{t}{x}{y}{t_1}) \cta \tau$}
  \DP\\[2ex]  
  


  \AXC{$\Gamma,x\cta\tau_1\vdash t\cta\tau_2$}
  \UIC{$\Gamma\vdash(\lambda x\cta\tau_1.t)\cta\tau_1\to\tau_2$}
  \DP
  \qquad\qquad
  \AXC{$\Gamma\vdash t_1\cta\tau_1\to\tau_2$}
  \AXC{$\Gamma\vdash t_2\cta\tau_1$}
  \BIC{$\Gamma\vdash (t_1~t_2)\cta\tau_2$}
  \DP\\[2ex]
  
  \AXC{$\Gamma \vdash t_1 \cta \tau_1 \to \tau_2 \to \tau_2$}
  \AXC{$\Gamma \vdash t_2 \cta \tau_2$}
  \AXC{$\Gamma \vdash t_3 \cta [\tau_1]$}
  \TIC{$\Gamma \vdash \lFold{t_1, t_2, t_3} \cta \tau_2$}
  \DP\\[2ex]

  \AXC{$\Gamma \vdash t_1 \cta \tau \to \tau$}
  \AXC{$\Gamma \vdash t_2 \cta \tau$}
  \AXC{$\Gamma \vdash t_3 \cta \Int$}
  \TIC{$\Gamma \vdash \iFold{t_1, t_2, t_3} \cta \tau$}
  \DP\\[1ex]

  \caption{Typing rules}
  \label{fig:typingrules}
\end{figure}

Semantically, types are interpreted as sets in an absolutely standard way, see Figure~\ref{fig:s:typesem} (where $\theta$ is a mapping from type variables to sets).
There is also a standard denotational term semantics, 
shown in Figure~\ref{fig:s:termsem}, 
which satisfies: if $\Gamma \vdash t \cta \tau$, then $\sem{t}_{\ts} \in \sem{\tau}_{\theta}$ for every $\sigma$ with $\sigma(x) \in \sem{\tau'}_{\theta}$ for every $x \cta \tau' $ in~$\Gamma$.%
\begin{figure}[p]
  \centering
  \[
  \begin{array}{@{}r@{\;}lr@{}}
    \sem{\alpha}_{\theta} & = {\theta(\alpha)} & \text{(an arbitrary set, fixed in $\theta$)} \\
    \sem{\Int}_{\theta}   & = {\N} & \text{(the naturals)}\\
    \sem{[\tau]}_{\theta} & \multicolumn{2}{@{}l@{}}{= {\{[\svv{x_1}, \ldots, \svv{x_n}] \mid n \in \N \wedge \forall i \in \{1,\ldots,n\}.~\svv{x_i} \in {\sem{\tau}_{\theta}}\}}\qquad\text{(the free monoid over a set)}}\\
    \sem{\tPair{\tau_1}{\tau_2}}_{\theta} & = {{\sem{\tau_1}_{\theta}}\times{\sem{\tau_2}_{\theta}}} & \text{(the Cartesian product of sets)}\\
    \sem{\tau_1 \to \tau_2}_{\theta} & = \fspace{\sem{\tau_1}_{\theta}}{\sem{\tau_2}_{\theta}} & \text{(the mathematical function space between sets)}
  \end{array}
  \]
  \caption{Standard type semantics}
  \label{fig:s:typesem}
\end{figure}%
\begin{figure}
  \centering
  \begin{align*}
    \sem{x}_{\ts} & = \sigma(x) \\
    \sem{n}_{\ts} & = \svv{n} \\
    \sem{\icase{t}{t_1}{x}{t_2}}_{\ts} & =
    \begin{cases}
      {\sem{t_1}_{\ts}} & \cif \sem{t}_{\ts} = \svv{0} \\
      {\sem{t_2}_{\ts[x \mapsto \svv{n}]}} & \cif \sem{t}_{\ts} = \svv{n},\; \svv{n} > \svv{0} 
    \end{cases} \\
    \sem{t_1 + t_2}_{\ts} & = 
    \sem{t_1}_{\ts} + \sem{t_2}_{\ts} \\
    \sem{\nil_{\tau}}_{\ts} & = \vnil \\
    \sem{t_1 : t_2}_{\ts} & = 
    [\sem{t_1}_{\ts},\svv{v_1},\ldots,\svv{v_n}] \with \sem{t_2}_{\ts} = [\svv{v_1},\ldots,\svv{v_n}]\\
    \sem{\lcase{t}{t_1}{x}{y}{t_2}}_{\ts} & =
    \begin{cases}
      {\sem{t_1}_{\ts}} & \cif \sem{t}_{\ts} = \vnil \\
      {\sem{t_2}_{\ts[x \mapsto \svv{v_1}, y \mapsto [\svv{v_2},\ldots,\svv{v_n}]]}} & \cif \sem{t}_{\ts} = [\svv{v_1},\ldots,\svv{v_n}],\; \svv{n} > \svv{0}
    \end{cases} \\
    \sem{(t_1, t_2)}_{\ts} & = (\sem{t_1}_{\ts},\sem{t_2}_{\ts}) \\
    \sem{\pcase{t}{x}{y}{t_1}}_{\ts} & =
    {\sem{t_1}_{\ts[x \mapsto \svv{v_1}, y \mapsto \svv{v_2}]}} \with \sem{t}_{\ts} = (\svv{v_1}, \svv{v_2}) \\
    \sem{\lambda x \cta \tau. t}_{\ts} & = \uplambda \svv{v}. {\sem{t}_{\ts[x \mapsto \svv{v}]}} \\
    \sem{t_1~t_2}_{\ts} & = \sem{t_1}_{\ts} ~ \sem{t_2}_{\ts} \\
    \sem{\lFold{t_1, t_2, t_3}}_{\ts} & = 
    {{\sem{t_1}_{\ts}~\svv{v_1} ~ (\sem{t_1}_{\ts}~\svv{v_2} ~\ldots (\sem{t_1}_{\ts}~\svv{v_n} ~ \sem{t_2}_{\ts})\ldots)}} 
    \with \sem{t_3}_{\ts} = [\svv{v_1}, \ldots, \svv{v_n}]
    \\
    \sem{\iFold{t_1, t_2, t_3}}_{\ts} & = 
    {{\underbrace{\sem{t_1}_{\ts} ~ (\sem{t_1}_{\ts} ~ \ldots (\sem{t_1}_{\ts} ~}_{\sem{t_3}_{\ts} \text{ times}}\: \sem{t_2}_{\ts})\ldots)}}
  \end{align*}%
  \caption{Standard term semantics}
  \label{fig:s:termsem}
\end{figure}%

The key to relational parametricity, and thus to free theorems, is to provide a suitable interpretation of types as relations.
The standard such type-indexed family of relations for our setting so far, defined by induction on the structure of types, and called a ``logical relation'', is given in Figure~\ref{fig:s:logrel} (where $\rho$ is a mapping from type variables to binary relations between sets).
Note that we use juxtaposition $(\svv{f}~\svv{x})$, instead of $\svv{f}(\svv{x})$, as notation for applying mathematical functions (mirroring the syntactic application on term level).
Also, we use the following definitions:
\[
\begin{array}{r@{\;}l}
  \llift{\rR} & = \{([\svv{x_1}, \ldots, \svv{x_n}], [\svv{y_1}, \ldots, \svv{y_n}]) \mid n \in \N \wedge \forall i \in \{1, \ldots, n\}.~ (\svv{x_i}, \svv{y_i}) \in \rR \} \\[1ex]
  \plift{\rR_1}{\rR_2} & =  \{ (\svv{(x_1, x_2)}, \svv{(y_1, y_2)}) \mid (\svv{x_1}, \svv{y_1}) \in \rR_1 \wedge (\svv{x_2}, \svv{y_2}) \in \rR_2 \}
\end{array}
\]%
\begin{figure}
\begin{align*}
  \slogrel{\alpha}{\rho}
  & = \rho(\alpha) \\
  \slogrel{\Int}{\rho}
  & = \id_{\N} \\
  \slogrel{[\tau]}{\rho} & = \llift{\slogrel{\tau}{\rho}}\\
  \slogrel{(\tau_1, \tau_2)} {\rho}
  & = \plift{\slogrel{\tau_1}{\rho}}{\slogrel{\tau_2}{\rho}} \\
  \slogrel{\tau_1 \to \tau_2}{\rho} & = \{ (\sv{f}, \sv{g}) \mid
  \forall (\svv{x}, \svv{y}) \in \slogrel{\tau_1}{\rho}.~
  (\svv{f}~\svv{x}, \svv{g}~\svv{y}) \in \slogrel{\tau_2}{\rho} \}
\end{align*}
\centering
\caption{Standard logical relation}
\label{fig:s:logrel}
\end{figure}


\pagebreak
To derive free theorems, all one needs is the following theorem~\citep{rey83,wad89}.
In it, $\rel$ denotes the collection of all binary relations between sets.
(Later, we also use $\rel(S_1, S_2)$ to denote more specifically the collection of all binary relations between sets $S_1$ and $S_2$.)
\begin{theorem}[standard parametricity theorem]\label{thm:s:para}
  If $\Gamma \vdash t \cta \tau$, then for every $\rho$, $\sigma_1$, $\sigma_2$ such that
  \begin{itemize}
  \item for every $\alpha$ in $\Gamma$, $\rho(\alpha) \in \rel$, and
  \item for every $x \cta \tau'$ in $\Gamma$, $(\sigma_1(x), \sigma_2(x)) \in \slogrel{\tau'}{\rho}$\,,
  \end{itemize}
  we have $(\sem{t}_{\tsone}, \sem{t}_{\tstwo}) \in \slogrel{\tau}{\rho}$.
\end{theorem}
Our aim now is to provide an analogous theorem for a setting in which computation costs are taken into account.
For doing so, we clearly first need to develop the underlying semantic notions (and then a suitable new logical relation).

\section{Adding costs to the semantics}

As already mentioned in the introduction, we want to study the call-by-value case here.
That is, we consider the presented lambda-calculus as a small core language of a kind of strict Haskell or pure ML.

In order to reflect computation costs in the semantics, we first revise the set interpretation of types.
In addition to a value, every semantic object now has to carry an integer representing some abstract notion of costs incurred while computing that value. 
Such integers (actually naturals would suffice for the moment, but the added generality of negative numbers comes in handy later on) need to be added only at top-level positions of compound values, thanks to our restriction to strict evaluation.
For example, the costs of individual list elements are not relevant ultimately, only the cost of a whole list, because anyway there is no means to evaluate only a part of it (as there would be in a nonstrict language).
The only place where ``embedded'' costs are relevant is in (the result positions of) function spaces, because there it is really important to capture which actual function arguments lead to which specific costs in the output.
Formally, we define a variant of the mapping from Figure~\ref{fig:s:typesem} in Figure~\ref{fig:typesem}, where 
$\Cost{S} = \{(\svv{x}, \svc{c}) \mid \svv{x} \in S\wedge \svc{c} \in \Z\}$.
That new mapping, $\ctsem{\cdot}$, does not itself capture top-level costs.
But ultimately, instead of the earlier $\sem{t}_{\ts} \in \sem{\tau}_{\theta}$ we will have that a term $t$ of type $\tau$ is mapped, by a new term semantics, to an element of the \emph{$\Cost{\cdot}$-lifting} of $\ctsem{\tau}_{\theta}$.
\begin{figure}[p]
  \begin{align*}
    \ctsem{\alpha}_{\theta} & = {\theta(\alpha)} \\
    \ctsem{\Int}_{\theta}   & = {\N} \\
    \ctsem{[\tau]}_{\theta} & = {\{[\svv{x_1}, \ldots, \svv{x_n}] \mid n \in \N \wedge \forall i \in \{1,\ldots,n\}.~ \svv{x_i} \in {\ctsem{\tau}_{\theta}}\}} \\
    \ctsem{\tPair{\tau_1}{\tau_2}}_{\theta} & = {{\ctsem{\tau_1}_{\theta}}\times {\ctsem{\tau_2}_{\theta}}} \\
    \ctsem{\tau_1 \to \tau_2}_{\theta} & = \fspace{\ctsem{\tau_1}_{\theta}}{\Cost{\ctsem{\tau_2}_{\theta}}}
  \end{align*}
  \centering
  \caption{Type semantics with embedded costs}
  \label{fig:typesem}
\end{figure}
 
Our new term semantics (changed from Figure~\ref{fig:s:termsem}) follows the same spirit as the instrumented semantics of \citet{ros89}.
Essentially, the cost integers are carried around and just suitably propagated, except where we decide that a certain semantic operation should be counted as contributing a cost of its own.
Here we assign a cost only to the invocation of functions, so we add a cost of $1$ in the interpretation of lambda-abstractions.\footnote{Other possible places to put extra costs would have been the data constructors and \textsyntax{case}-expressions. Actually, we have found that our general results, in particular Theorem~\ref{thm:para}, are unaffected by such changes.}
The formal definition is given in Figure~\ref{fig:costtermsem}. 
The helper function $\addCost{}{}$ defined in the figure adds, in $\addCost{\svc{c}}{\sv{x}}$, the cost $\svc{c}$ to the cost component of semantic object $\sv{x}$.
The other helper functions are cost-propagating versions of data constructors and function application.
Syntactically, $\addCost{}{}$ and $\cOpCons$ are right-associative, $\cOpApp$ is left-associative, and 
$\addCost{}{}$ has higher precedence than the other semantic operations.
Now we have that if $\Gamma \vdash t \cta \tau$ then $\csem{t}_{\ts} \in \Cost{\ctsem{\tau}_{\theta}}$ for every $\theta$ mapping the type variables in $\Gamma$ to sets and $\sigma$ with $\sigma(x) \in \ctsem{\tau'}_{\theta}$ for every $x \cta \tau'$ in $\Gamma$.
\begin{figure}
  \begin{align*}
    \csem{x}_{\ts} & = (\sigma(x), \svc{0}) \\
    \csem{n}_{\ts} & = (\svv{n}, \svc{0}) \\
    \csem{\icase{t}{t_1}{x}{t_2}}_{\ts} & =
      \begin{cases}
        \addCost{\svc{c}}{\csem{t_1}_{\ts}} & \cif \csem{t}_{\ts} = (\svv{0}, \svc{c}) \\
        \addCost{\svc{c}}{\csem{t_2}_{\ts[x \mapsto \svv{n}]}} & \cif \csem{t}_{\ts} = (\svv{n}, \svc{c}),\; \svv{n} > \svv{0} 
      \end{cases} \\
    \csem{t_1 + t_2}_{\ts} & = 
        (\svv{n_1} + \svv{n_2}, \svc{c_1} + \svc{c_2}) \with \csem{t_1}_{\ts} = (\svv{n_1}, \svc{c_1}),\; \csem{t_2}_{\ts} = (\svv{n_2}, \svc{c_2}) \\
    \csem{\nil_{\tau}}_{\ts} & = (\vnil, \svc{0}) \\
    \csem{t_1 : t_2}_{\ts} & = 
        \csem{t_1}_{\ts} \cOpCons \csem{t_2}_{\ts} \\
    \csem{\lcase{t}{t_1}{x}{y}{t_2}}_{\ts} & =
      \begin{cases}
        \addCost{\svc{c}}{\csem{t_1}_{\ts}} & \cif \csem{t}_{\ts} = (\vnil, \svc{c}) \\
        \addCost{\svc{c}}{\csem{t_2}_{\ts[x \mapsto \svv{v_1}, y \mapsto [\svv{v_2},\ldots,\svv{v_n}]]}} & \cif \csem{t}_{\ts} = ([\svv{v_1},\ldots,\svv{v_n}], \svc{c}),\; \svv{n} > \svv{0}
      \end{cases} \\
    \csem{(t_1, t_2)}_{\ts} & = \cOpPair{\csem{t_1}_{\ts}}{\csem{t_2}_{\ts}} \\
    \csem{\pcase{t}{x}{y}{t_1}}_{\ts} & =
        \addCost{\svc{c}}{\csem{t_1}_{\ts[x \mapsto \svv{v_1}, y \mapsto \svv{v_2}]}} \with \csem{t}_{\ts} = ((\svv{v_1}, \svv{v_2}), \svc{c}) \\
    \csem{\lambda x \cta \tau. t}_{\ts} & = (\uplambda \svv{v}. \addCost{\cApp}{\csem{t}_{\ts[x \mapsto \svv{v}]}}, \svc{0}) \\
    \csem{t_1~t_2}_{\ts} & = \csem{t_1}_{\ts} \cOpApp \csem{t_2}_{\ts} \\
    \csem{\lFold{t_1, t_2, t_3}}_{\ts} & = 
        {\addCost{(\svc{c_1}+\svc{c_3})}{((\svv{g}~\svv{v_1}) \cOpApp ((\svv{g}~\svv{v_2}) \cOpApp \ldots ((\svv{g}~\svv{v_n}) \cOpApp \csem{t_2}_{\ts})\ldots))}} \\
        &\quad \with  \csem{t_1}_{\ts} = (\svv{g}, \svc{c _1}),\; \csem{t_3}_{\ts} = ([\svv{v_1}, \ldots, \svv{v_n}], \svc{c_3})
        \\
    \csem{\iFold{t_1, t_2, t_3}}_{\ts} & = 
    {\addCost{(\svc{c_1}+\svc{c_3})}{(\underbrace{(\svv{g}, \svc{0}) \cOpApp ((\svv{g}, \svc{0}) \cOpApp\ldots ((\svv{g}, \svc{0}) \cOpApp}_{\svv{n}\text{ times}}\: \csem{t_2}_{\ts})\ldots))}} \\
    & \quad \with \csem{t_1}_{\ts} = (\svv{g}, \svc{c_1}),\; \csem{t_3}_{\ts} = (\svv{n}, \svc{c_3})
    \\
  \end{align*}
  where
  \[
  \begin{array}{c@{\;}l}
    \addCost{\svc{c}}{(\sv{v}, \svc{c'})} & = (\sv{v}, \svc{c} + \svc{c'}) \\[1ex]
    \sv{x} \cOpCons \sv{xs}
    & = ([\svv{v},\svv{v_1},\ldots,\svv{v_n}], \svc{c} + \svc{c'}) \with \sv{x} = (\svv{v}, \svc{c}),\;  \sv{xs} = ([\svv{v_1},\ldots,\svv{v_n}], \svc{c'}) \\[1ex]
    \cOpPair{\sv{x_1}}{\sv{x_2}}
    & = ((\svv{v_1}, \svv{v_2}), \svc{c} + \svc{c'}) \with \sv{x_1} = (\svv{v_1}, \svc{c}),\; \sv{x_2} = (\svv{v_2}, \svc{c'}) \\[1ex]
    \sv{f} \cOpApp \sv{x} 
    & = 
        {\addCost{(\svc{c}+\svc{c'})}{(\sv{g}~\sv{v})}} \with \sv{f} = (\svv{g}, \svc{c}),\; \sv{x} = (\svv{v}, \svc{c'}) 
  \end{array}
  \]
  \centering
  \caption{Term semantics with costs
}
  \label{fig:costtermsem}
\end{figure}
\begin{example}
  Let $\var{length}=\lambda \xs\cta[\alpha].\lFold{\lambda x\cta\alpha. \lambda y\cta\Int. 1 + y, 0, \xs}$.
  We calculate the semantics of $\var{length}[\Int/\alpha]~(1:2:[\,]_\Int)$, where $[\Int/\alpha]$ denotes syntactic substitution of $\Int$ for all occurrences of $\alpha$, as follows:
  \[
  \begin{array}{@{}l@{\;}l@{}}
    & \csem{(\lambda \xs\cta[\Int].\lFold{\lambda x\cta\Int. \lambda y\cta\Int. 1 + y, 0, \xs})~(1:2:[\,]_\Int)}_\ets\\
    = & (\uplambda \svv{v}. \addCost{\cApp}{\csem{\lFold{\lambda x\cta\Int. \lambda y\cta\Int. 1 + y, 0, \xs}}_{[\xs \mapsto \svv{v}]}}, \svc{0}) \cOpApp (\svv{[1, 2]}, \svc{0})\\
    = & \addCost{\cApp}{\csem{\lFold{\lambda x\cta\Int. \lambda y\cta\Int. 1 + y, 0, \xs}}_{[\xs \mapsto \svv{[1, 2]}]}}\\
    = & \addCost{\cApp}{(((\uplambda \svv{x}. (\uplambda \svv{y}. (\svv{1} + \svv{y}, \cApp), \cApp))~\svv{1}) \cOpApp (((\uplambda \svv{x}. (\uplambda \svv{y}. (\svv{1} + \svv{y}, \cApp), \cApp))~\svv{2}) \cOpApp (\svv{0}, \svc{0})))}\\
    = & \addCost{\cApp}{((\uplambda \svv{y}. (\svv{1} + \svv{y}, \cApp), \cApp) \cOpApp \addCost{\cApp}{(\svv{1} + \svv{0}, \cApp)})}\\
    = & \addCost{\cApp}{
      {\addCost{(\cApp+\cApp+\cApp)}{((\uplambda \svv{y}. (\svv{1} + \svv{y}, \cApp))~(\svv{1} + \svv{0}))}}}\\
    = & (\svv{2}, 5)
  \end{array}
  \]
  Exactly the five required beta-reductions (once for $\lambda \xs\cta[\Int]$ and twice each for $\lambda x\cta\Int. \lambda y\cta\Int$) have been counted.
\end{example}

Note that due to the way we handle polymorphism, a $\csem{t}_{\ts}$ can be element of $\Cost{\ctsem{\tau}_{\theta_1}}$ and $\Cost{\ctsem{\tau}_{\theta_2}}$ for completely different $\theta_1$ and $\theta_2$.
For example, $\csem{(\lambda x\cta\alpha.x)}_{\ets}$ is $(\sv{g},0)$ where $\sv{g}$ maps $\sv{v}\in S$ to $(\sv{v},1)\in\Cost{S}$, for \emph{every} set $S$.
(We denote by $\ets$ an empty mapping.)
\begin{lemma}\label{lem:typeinst}
  Let $\Gamma \vdash t \cta \tau$, where $\Gamma$ contains no term variables.
  For every type variable $\alpha$, type $\tau'$ not containing type variables, and $\theta$ mapping the type variables in $\Gamma\setminus\{\alpha\}$ to sets, we have $\csem{t[\tau'/\alpha]}_{\ets} \in \Cost{\ctsem{\tau[\tau'/\alpha]}_{\theta}}$.
  Moreover, $\ctsem{\tau[\tau'/\alpha]}_{\theta}=\ctsem{\tau}_{\theta[\alpha\mapsto\ctsem{\tau'}_\ets]}$, and while $\csem{t}_{\ets}$ is an element of $\Cost{\ctsem{\tau}_{\theta[\alpha\mapsto S]}}$ for \emph{arbitrary} $S$, for the \emph{specific} case $S=\ctsem{\tau'}_\ets$ we have $\csem{t}_{\ets}=\csem{t[\tau'/\alpha]}_{\ets}$.
\end{lemma}

We also note some simple properties of the semantic operations; these properties will henceforth be used freely without explicit mention:\\
\parbox{0.5\textwidth}{
\begin{itemize}

\item 
  $\addCost{\svc{c}}{\addCost{\svc{c'}}{\sv{x}}} = \addCost{(\svc{c} + \svc{c'})}{\sv{x}}$

\item 
  $\addCost{\svc{c}}{(\sv{x} \cOpCons \sv{xs})} = \addCost{\svc{c}}{\sv{x}} \cOpCons \sv{xs} = \sv{x} \cOpCons  \addCost{\svc{c}}{\sv{xs}}$
\end{itemize}}
\parbox{0.5\textwidth}{
\begin{itemize}
\item 
  $\addCost{\svc{c}}{\cOpPair{\sv{x_1}}{\sv{x_2}}} = \cOpPair{\addCost{\svc{c}}{\sv{x_1}}}{\sv{x_2}} = \cOpPair{\sv{x_1}}{\addCost{\svc{c}}{\sv{x_2}}}$

\item 
  $\addCost{\svc{c}}{(\sv{f} \cOpApp \sv{x})} = \addCost{\svc{c}}{\sv{f}} \cOpApp \sv{x} = \sv{f} \cOpApp \addCost{\svc{c}}{\sv{x}}$
\end{itemize}
}

\section{New relational interpretations of types}

Now we also need a new interpretation of types as relations, i.e., a new logical relation.
We get directions by comparing the set interpretations from Figures~\ref{fig:s:typesem} and~\ref{fig:typesem}.
There, a difference only appears for the output side of function arrows, namely the codomain is lifted to a costful setting.
We try the same on the relational level and thus transform the logical relation from Figure~\ref{fig:s:logrel} into the one given in Figure~\ref{fig:logrel}, where $\Cost{\rR} = \{((\svv{x}, \svc{c}), (\svv{y}, \svc{c})) \mid (\svv{x}, \svv{y}) \in \rR \wedge \svc{c} \in \Z \}$.
\begin{figure}
\begin{align*}
  \logrel{\alpha}{\rho}
  & = \rho(\alpha) \\
  \logrel{\Int}{\rho}
  & = \id_{\N} \\
  \logrel{[\tau]}{\rho} & = \llift{\logrel{\tau}{\rho}} \\
  \logrel{(\tau_1, \tau_2)} {\rho}
  & = \plift{\logrel{\tau_1}{\rho}}{\logrel{\tau_2}{\rho}} \\
  \logrel{\tau_1 \to \tau_2}{\rho} & = \{ (\sv{f}, \sv{g}) \mid
  \forall (\svv{x}, \svv{y}) \in \logrel{\tau_1}{\rho}.~
  (\svv{f}~\svv{x}, \svv{g}~\svv{y}) \in \Cost{\logrel{\tau_2}{\rho}} \}
\end{align*}
\centering
\caption{Logical relation with embedded costs}
\label{fig:logrel}
\end{figure}


Note that $\rho$ in Figure~\ref{fig:logrel} still maps to ``normal'' binary relations between sets, rather than to $\Cost{\cdot}$-lifted ones.
In turn, the $\csem{\cdot}$-semantics of terms will be related by the \emph{$\Cost{\cdot}$-lifting} of $\plogrel$.
Indeed, a proof very similar to that of Theorem~\ref{thm:s:para}, by induction on typing derivations, establishes the following theorem.
(The proof is sketched in Appendix~\ref{app:para}.)
\begin{theorem}\label{thm:para}
  If $\Gamma \vdash t \cta \tau$, then for every $\rho$, $\sigma_1$, $\sigma_2$ such that
  \begin{itemize}
  \item for every $\alpha$ in $\Gamma$, $\rho(\alpha) \in \rel$, and
  \item for every $x \cta \tau'$ in $\Gamma$, $(\sigma_1(x), \sigma_2(x)) \in \logrel{\tau'}{\rho}$\,,
  \end{itemize}
  we have $(\csem{t}_{\tsone}, \csem{t}_{\tstwo}) \in \Cost{\logrel{\tau}{\rho}}$.
\end{theorem}
One of the key cases in the proof, for function application, uses that $(\sv{f}, \sv{g}) \in \Cost{\logrel{\tau_1 \to \tau_2}{\rho}}$ implies $\forall (\sv{x}, \sv{y}) \in \Cost{\logrel{\tau_1}{\rho}}.~(\sv{f} \cOpApp \sv{x}, \sv{g} \cOpApp \sv{y}) \in \Cost{\logrel{\tau_2}{\rho}}$.
Note the subtle differences here to the definition of $\logrel{\tau_1 \to \tau_2}{\rho}$ in Figure~\ref{fig:logrel}, namely the $\Cost{\cdot}$-lifting on both $\logrel{\tau_1 \to \tau_2}{\rho}$ and $\logrel{\tau_1}{\rho}$, and hence the use of $(\sv{f} \cOpApp \sv{x},\sv{g} \cOpApp \sv{y})$ instead of $(\svv{f}~\svv{x},\svv{g}~\svv{y})$.
Working fully on the $\Cost{\cdot}$-lifted level is also preferable in later derivations of free theorems (based on the logical relation), so it seems a good idea to provide an alternative definition of relational interpretations of types that does not mix unlifted (like $\logrel{\tau_1}{\rho}$) and lifted (like $\Cost{\logrel{\tau_2}{\rho}}$) uses.
However, we have to be careful, because the ``implies'' in the first sentence of the current paragraph is really just that: an implication, not an equivalence.
In order to give a direct inductive definition for $\Cost{\logrel{\,\cdot}{\cdot}}$, we need exact characterizations.
For the case of function types, the following lemma is easily obtained from the definitions
, where, in general, $\getCost{(\svv{v}, \svc{c})} = \svc{c}$.
\begin{lemma}\label{lem:lrfun}
  $(\sv{f}, \sv{g}) \in \Cost{\logrel{\tau_1 \to \tau_2}{\rho}} \eqa \getCost{\sv{f}} = \getCost{\sv{g}}
  \wedge \forall (\sv{x}, \sv{y}) \in \Cost{\logrel{\tau_1}{\rho}}.~
  (\sv{f} \cOpApp \sv{x}, \sv{g} \cOpApp \sv{y}) \in \Cost{\logrel{\tau_2}{\rho}}
  $
\end{lemma}
Using similar characterizations for the other cases, we arrive at the new logical relation given in Figure~\ref{fig:clogrel}, which is connected to the one from Figure~\ref{fig:logrel} by the following (inductively proved) lemma.
\begin{lemma}
  For every $\tau$ and $\rho$,
  $\Cost{\logrel{\tau}{\rho}}=\clogrel{\tau}{\rho}$.
\end{lemma}
\begin{figure}
\begin{align*}
  \clogrel{\alpha}{\rho}
  & = \Cost{\rho(\alpha)} \\
  \clogrel{\Int}{\rho}
  & = \id_{\Cost{\N}} \\
  \clogrel{[\tau]}{\rho} & = \cllift{\clogrel{\tau}{\rho}}\\
  \clogrel{(\tau_1, \tau_2)} {\rho}
  & = \cplift{\clogrel{\tau_1}{\rho}}{\clogrel{\tau_2}{\rho}} \\
  \clogrel{\tau_1 \to \tau_2}{\rho} & = \{(\sv{f}, \sv{g}) \mid  \getCost{\sv{f}} = \getCost{\sv{g}}
  \wedge \forall (\sv{x}, \sv{y}) \in \clogrel{\tau_1}{\rho}.~ 
  (\sv{f} \cOpApp \sv{x}, \sv{g} \cOpApp \sv{y}) \in \clogrel{\tau_2}{\rho}
  \}
\end{align*}
where
\[
\begin{array}{r@{\;}l}
  \cllift{\rR^\text{\rm{\textcent}}} &= \{(\clist{\sv{x_1}, \ldots, \sv{x_n}}, \clist{\sv{y_1}, \ldots, \sv{y_n}}) \mid n \in \N \wedge \forall i \in \{1, \ldots, n\}.~(\sv{x_i}, \sv{y_i}) \in \rR^\text{\rm{\textcent}}\}\\[1ex]
  \cplift{\rR^\text{\rm{\textcent}}_1}{\rR^\text{\rm{\textcent}}_2} &= \{(\cOpPair{\sv{x_1}}{\sv{x_2}}, \cOpPair{\sv{y_1}}{\sv{y_2}}) \mid (\sv{x_1}, \sv{y_1}) \in \rR^\text{\rm{\textcent}}_1 \wedge (\sv{x_2}, \sv{y_2}) \in \rR^\text{\rm{\textcent}}_2 \}
\end{array}
\]
and $\clist{\sv{x_1}, \ldots, \sv{x_n}}$ abbreviates $\sv{x_1} \cOpCons \ldots \cOpCons \sv{x_n} \cOpCons (\vnil, \svc{0})$.
\caption{Fully cost-lifted logical relation}
\label{fig:clogrel}
\end{figure}

Together with Theorem~\ref{thm:para}, we immediately get:
\begin{corollary} \label{cor:para}
  If $\Gamma \vdash t \cta \tau$, then for every $\rho$, $\sigma_1$, $\sigma_2$ such that
  \begin{itemize}
  \item for every $\alpha$ in $\Gamma$, $\rho(\alpha) \in \rel$, and
  \item for every $x \cta \tau'$ in $\Gamma$, $((\sigma_1(x),0), (\sigma_2(x),0)) \in \clogrel{\tau'}{\rho}$\,,
  \end{itemize}
  we have $(\csem{t}_{\tsone}, \csem{t}_{\tstwo}) \in \clogrel{\tau}{\rho}$.
\end{corollary}

\section{Deriving free theorems}

Now we can go for applications of Corollary~\ref{cor:para} to specific polymorphic types, in order to derive cost-aware statements about terms of those types.
First, we need some auxiliary notions.
In addition to $\getCost{(\svv{v}, \svc{c})} = \svc{c}$ we define $\getVal{(\svv{v}, \svc{c})} = \sv{v}$, and for every $\sv{f}\in\Cost{\fspace{S_1}{\Cost{S_2}}}$ and $\sv{x}\in\Cost{S_1}$, for some sets $S_1$ and $S_2$,
$\appCost(\sv{f}, \sv{x}) = \getCost{\sv{f} \cOpApp \sv{x}} - \getCost{\sv{x}}$.
Also, a standard way of deriving free theorems is to specialize relations (those mapped to by $\rho$) to the graphs of functions.
In our setting, we have to be careful to get the ``$\Cost{\cdot}$-lifting level'' right.
Moreover, since in our derivations of free theorems we will need to have access to information about the costs associated to specific function arguments and results, it is helpful to make specialized relations as tightly specified as possible.
Hence, instead of the full function graphs commonly used, we will go for finite parts thereof.
So given sets $S_1$ and $S_2$, a $\Cost{\cdot}$-lifted function $\sv{g} \in \Cost{\fspace{S_1}{\Cost{S_2}}}$, and $\Cost{\cdot}$-lifted values $\sv{x_1}, \ldots, \sv{x_n} \in \Cost{S_1}$, with $n \in \N$, we define:
\[\rR^{\sv{g}}_{\sv{x_1}, \ldots, \sv{x_n}} = \{ (\getVal{\sv{x_1}}, \getVal{\sv{g} \cOpApp \sv{x_1}}), \ldots, (\getVal{\sv{x_n}}, \getVal{\sv{g} \cOpApp \sv{x_n}})\} \in \rel(S_1, S_2)\]
The crucial property, directly derived from definitions, (and a simple corollary of it) we are going to exploit about $\rR^{\sv{g}}_{\sv{x_1}, \ldots, \sv{x_n}}$ can be given as follows (under the given conditions on $S_1$, $S_2$, $\sv{g}$, and $\sv{x_1}, \ldots, \sv{x_n}$):
\begin{proposition}\label{prop:relspec}
  Let 
  $\sv{x}\in \Cost{S_1}$ and $\sv{y}\in \Cost{S_2}$.
  Then $(\sv{x},\sv{y})\in\Cost{\rR^{\sv{g}}_{\sv{x_1}, \ldots, \sv{x_n}}}$ if and only if there exist $i\in\{1,\ldots,n\}$ and $c\in\Z$ such that $\sv{x}=\addCost{c}{\addCost{\appCost(\sv{g},\sv{x_i})}{\sv{x_i}}}$ and $\sv{y}=\addCost{c}{(\sv{g} \cOpApp \sv{x_i})}$.
\end{proposition}
\begin{corollary}\label{cor:relspec}
  Let 
  $\sv{x}\in \Cost{S_1}$ and $\sv{y}\in \Cost{S_2}$.
  If $(\sv{x},\sv{y})\in\Cost{\rR^{\sv{g}}_{\sv{x_1}, \ldots, \sv{x_n}}}$, then there exists $i\in\{1,\ldots,n\}$ such that $\sv{g} \cOpApp \sv{x}=\addCost{\appCost(\sv{g},\sv{x_i})}{\sv{y}}$.
\end{corollary}

Let us now derive a first concrete free (improvement) theorem, for one of the 
types from Section~\ref{sec:intro}
.
\begin{example}
  Let some term $f$ be given with $\alpha \vdash f \cta \alpha \to \alpha \to \alpha$.
  By Corollary~\ref{cor:para} we have:
  \[\forall \rR \in \rel.~(\csem{f}_{\ets}, \csem{f}_{\ets}) \in \clogrel{\alpha \to \alpha \to \alpha}{[\alpha \mapsto \rR]}\]
  By the definition of the logical relation in Figure~\ref{fig:clogrel} this implies:
  \[\forall \rR \in \rel,(\sv{x}, \svv{y}),(\sv{x'}, \sv{y'}) \in \Cost{\rR}.~(\csem{f}_{\ets} \cOpApp \sv{x} \cOpApp \sv{x'}, \csem{f}_{\ets} \cOpApp \sv{y} \cOpApp \sv{y'}) \in  \Cost{\rR}\]
  Specialization of $\rR$ gives:
  \[
  \begin{array}{@{}l@{}}
    \forall S_1, S_2\text{ sets}, \sv{g} \in \Cost{\fspace{S_1}{\Cost{S_2}}}, \sv{x_1}, \sv{x_2} \in \Cost{S_1}.\\
    ~\forall (\sv{x}, \svv{y}),(\sv{x'}, \sv{y'}) \in \Cost{\rR^{\sv{g}}_{\sv{x_1}, \sv{x_2}}}.~(\csem{f}_{\ets} \cOpApp \sv{x} \cOpApp \sv{x'}, \csem{f}_{\ets} \cOpApp \sv{y} \cOpApp \sv{y'}) \in  \Cost{\rR^{\sv{g}}_{\sv{x_1}, \sv{x_2}}}
  \end{array}
  \]
  From this follows, by Proposition~\ref{prop:relspec}:
  \[
  \begin{array}{@{}l@{}}
    \forall S_1, S_2\text{ sets}, \sv{g} \in \Cost{\fspace{S_1}{\Cost{S_2}}}, \sv{x_1}, \sv{x_2} \in \Cost{S_1}.\\
    ~(\csem{f}_{\ets} \cOpApp (\addCost{\appCost(\sv{g}, \sv{x_1})}{\sv{x_1}}) \cOpApp (\addCost{\appCost(\sv{g}, \sv{x_2})}{\sv{x_2}}), \csem{f}_{\ets} \cOpApp (\sv{g} \cOpApp \sv{x_1}) \cOpApp (\sv{g} \cOpApp \sv{x_2})) \in  \Cost{\rR^{\sv{g}}_{\sv{x_1}, \sv{x_2}}}
  \end{array}
  \]
  which in turn implies, by Corollary~\ref{cor:relspec}:
  \[
  \begin{array}{@{}l@{}}
    \forall S_1, S_2\text{ sets}, \sv{g} \in \Cost{\fspace{S_1}{\Cost{S_2}}}, \sv{x_1}, \sv{x_2} \in \Cost{S_1}.~\exists i\in\{1,2\}.\\
    ~
    \begin{array}[t]{@{}l@{}}
      \sv{g} \cOpApp (\csem{f}_{\ets} \cOpApp (\addCost{\appCost(\sv{g}, \sv{x_1})}{\sv{x_1}}) \cOpApp (\addCost{\appCost(\sv{g}, \sv{x_2})}{\sv{x_2}}))\\
      = \addCost{\appCost(\sv{g},\sv{x_i})}{(\csem{f}_{\ets} \cOpApp (\sv{g} \cOpApp \sv{x_1}) \cOpApp (\sv{g} \cOpApp \sv{x_2}))}
    \end{array}
  \end{array}
  \]
  which simplifies to:
  \[
  \begin{array}{@{}l@{}}
    \forall S_1, S_2\text{ sets}, \sv{g} \in \Cost{\fspace{S_1}{\Cost{S_2}}}, \sv{x_1}, \sv{x_2} \in \Cost{S_1}.~\exists c\in\{\appCost(\sv{g},\sv{x_1}),\appCost(\sv{g},\sv{x_2})\}.\\
    ~\addCost{c}{(\sv{g} \cOpApp (\csem{f}_{\ets} \cOpApp \sv{x_1} \cOpApp \sv{x_2}))} = \csem{f}_{\ets} \cOpApp (\sv{g} \cOpApp \sv{x_1}) \cOpApp (\sv{g} \cOpApp \sv{x_2})
  \end{array}
  \]
  By using the definitions from Figures~\ref{fig:typesem} and~\ref{fig:costtermsem}, and Lemma~\ref{lem:typeinst}, we can conclude that:
  \[
  \begin{array}{@{}l@{}}
    \forall \tau_1, \tau_2\text{ types}, g\cta \tau_1 \to \tau_2, t_1\cta\tau_1, t_2\cta\tau_1.~\exists c\in\{\appCost(\csem{g}_{\ets},\csem{t_1}_{\ets}),\appCost(\csem{g}_{\ets},\csem{t_2}_{\ets})\}.\\
    ~\addCost{c}{\csem{g~(f[\tau_1/\alpha]~t_1~t_2)}_{\ets}} = \csem{f[\tau_2/\alpha]~(g~t_1)~(g~t_2)}_{\ets}
  \end{array}
  \]
  This certainly means that the right-hand side of~(\ref{eq:proj}) in the introduction is more efficient than its left-hand side.
  Indeed, after defining ``$\sv{v} \sqsubseteq \sv{v'}$'' as ``$\exists c\geq0.~\addCost{c}{\sv{v}} = \sv{v'}$'' (or, equivalently, ``$\getVal{\sv{v}}=\getVal{\sv{v'}} \wedge \getCost{\sv{v}} \leq \getCost{\sv{v'}}$''), we can conclude from the above that:
  \[
    \forall \tau_1, \tau_2\text{ types}, g\cta \tau_1 \to \tau_2, t_1\cta\tau_1, t_2\cta\tau_1.
    ~\csem{g~(f[\tau_1/\alpha]~t_1~t_2)}_{\ets} \sqsubseteq \csem{f[\tau_2/\alpha]~(g~t_1)~(g~t_2)}_{\ets}
  \]
  In the interest of readability, we will sometimes blur the distinction between syntax and semantics a bit, and additionally keep type substitution (for instantiating polymorphic functions) silent, so that the above conclusion would be written as simply
  \begin{equation}
    \label{eq:c:proj}
    g~(f~t_1~t_2) \sqsubseteq f~(g~t_1)~(g~t_2)
  \end{equation}

  To emphasize again that we crucially exploit polymorphism, recall from the introduction that a corresponding statement does \emph{not} hold for $f :: \Int\to\Int\to\Int$.
  Even if $\sem{f}_{\ets}=\sem{\lambda x\cta\Int.\lambda y\cta\Int.y}_{\ets}$ in the cost-free semantics, there can be $g::\Int\to\Int$ and $t_1,t_2::\Int$ such that, of course, $\getVal{\csem{g~(f~t_1~t_2)}_{\ets}} = \getVal{\csem{f~(g~t_1)~(g~t_2)}_{\ets}}$ is true, but~(\ref{eq:c:proj}) is false.
\end{example}

Let us now move on to other examples, like~(\ref{eq:pair}) in the introduction.
First, we define $\vmapPair=\csem{\mapPair}_{\ets}$ for some reasonable rendering of the $\mapPair$-function in our calculus.
Then, we can give analogues of Proposition~\ref{prop:relspec} and Corollary~\ref{cor:relspec} for pair-lifting, given sets $S_1$, $S_2$, $S_3$, and $S_4$, $\Cost{\cdot}$-lifted functions $\sv{g} \in \Cost{\fspace{S_1}{\Cost{S_2}}}$ and $\sv{h} \in \Cost{\fspace{S_3}{\Cost{S_4}}}$, and $\Cost{\cdot}$-lifted values $\sv{x_1}, \ldots, \sv{x_n} \in \Cost{S_1}$ and $\sv{y_1}, \ldots, \sv{y_m} \in \Cost{S_3}$, with $n,m \in \N$.
\begin{proposition}\label{prop:relspec:pair}
  Let $\sv{p}\in \Cost{S_1\times S_3}$ and $\sv{q}\in \Cost{S_2\times S_4}$.
  Then $(\sv{p}, \sv{q}) \in \cplift{\Cost{\rR^{\sv{g}}_{\sv{x_1}, \ldots, \sv{x_n}}}}{\Cost{\rR^{\sv{h}}_{\sv{y_1}, \ldots, \sv{y_m}}}}$ if and only if there exist $i \in \{1,\ldots, n\}$, $j \in \{1, \ldots m\}$, and $\svc{c} \in \Z$ such that $\sv{p} = \addCost{\svc{c}}{\addCost{\appCost(\vmapPair \cOpApp \cOpPair{\sv{g}}{\sv{h}}, \cOpPair{\sv{x_i}}{\sv{y_j}})}{\cOpPair{\sv{x_i}}{\sv{y_j}}}}$ and $\sv{q} = \addCost{\svc{c}}{(\vmapPair \cOpApp \cOpPair{\sv{g}}{\sv{h}} \cOpApp \cOpPair{\sv{x_i}}{\sv{y_j}})}$.
\end{proposition}
\begin{corollary}\label{cor:relspec:pair}
  Let $\sv{p}\in \Cost{S_1\times S_3}$ and $\sv{q}\in \Cost{S_2\times S_4}$.
  If $(\sv{p}, \sv{q}) \in \cplift{\Cost{\rR^{\sv{g}}_{\sv{x_1}, \ldots, \sv{x_n}}}}{\Cost{\rR^{\sv{h}}_{\sv{y_1}, \ldots, \sv{y_m}}}}$, then there exist $i \in \{1, \ldots, n\}$ and $j \in \{1, \ldots, m\}$ such that $\vmapPair \cOpApp \cOpPair{\sv{g}}{\sv{h}} \cOpApp \sv{p} = \addCost{\appCost(\vmapPair \cOpApp \cOpPair{\sv{g}}{\sv{h}}, \cOpPair{\sv{x_i}}{\sv{y_j}})}{\sv{q}}$.
\end{corollary}

Now we can deal with example types involving pairs.
\begin{example}
  Let some term $f$ be given with $\alpha \vdash f \cta \alpha \to (\alpha, \alpha)$.
  By Corollary~\ref{cor:para} we have:
  \[\forall \rR \in \rel.~(\csem{f}_{\ets}, \csem{f}_{\ets}) \in \clogrel{\alpha \to (\alpha, \alpha)}{[\alpha \mapsto \rR]}\]
  By the definition of the logical relation and specialization of $\rR$, this gives:
  \[
  \begin{array}{@{}l@{}}
    \forall S_1, S_2\text{ sets}, \sv{g} \in \Cost{\fspace{S_1}{\Cost{S_2}}}, \sv{x_1} \in \Cost{S_1}.\\
    ~\forall (\sv{x}, \svv{y}) \in \Cost{\rR^{\sv{g}}_{\sv{x_1}}}.~(\csem{f}_{\ets} \cOpApp \sv{x}, \csem{f}_{\ets} \cOpApp \sv{y}) \in  \cplift{\Cost{\rR^{\sv{g}}_{\sv{x_1}}}}{\Cost{\rR^{\sv{g}}_{\sv{x_1}}}}
  \end{array}
  \]
  From this follows, by Proposition~\ref{prop:relspec} 
  and Corollary~\ref{cor:relspec:pair}:
  \[
  \begin{array}{@{}l@{}}
    \forall S_1, S_2\text{ sets}, \sv{g} \in \Cost{\fspace{S_1}{\Cost{S_2}}}, \sv{x_1} \in \Cost{S_1}.\\
    ~
    \begin{array}[t]{@{}l@{}}
      \vmapPair \cOpApp \cOpPair{\sv{g}}{\sv{g}} \cOpApp (\csem{f}_{\ets} \cOpApp (\addCost{\appCost(\sv{g}, \sv{x_1})}{\sv{x_1}}))\\
      = \addCost{\appCost(\vmapPair \cOpApp \cOpPair{\sv{g}}{\sv{g}}, \cOpPair{\sv{x_1}}{\sv{x_1}})}{(\csem{f}_{\ets} \cOpApp (\sv{g} \cOpApp \sv{x_1}))}
    \end{array}
  \end{array}
  \]
  which due to the certainly nonnegative difference $\appCost(\vmapPair \cOpApp \cOpPair{\sv{g}}{\sv{g}}, \cOpPair{\sv{x_1}}{\sv{x_1}})-\appCost(\sv{g}, \sv{x_1})$ simplifies to:
  \[\forall \tau_1, \tau_2\text{ types}, g\cta \tau_1 \to \tau_2, t\cta\tau_1.~f~(g~t)\sqsubseteq\mapPair~(g, g)~(f~t)\]
\end{example}
\begin{example}
  Let some term $f$ be given with $\alpha \vdash f \cta (\alpha, \alpha) \to \alpha$.
  Using Corollary~\ref{cor:para}, the definition of the logical relation, and Proposition~\ref{prop:relspec:pair} and Corollary~\ref{cor:relspec} for $\rR^{\sv{g}}_{\sv{x_1}, \sv{x_2}}$, we get:
  \[
  \begin{array}{@{}l@{}}
    \forall S_1, S_2\text{ sets}, \sv{g} \in \Cost{\fspace{S_1}{\Cost{S_2}}}, \sv{x_1},\sv{x_2} \in \Cost{S_1}.~\exists c\in\{\appCost(\sv{g}, \sv{x_1}),\appCost(\sv{g}, \sv{x_2})\}.\\
    ~
    \begin{array}[t]{@{}l@{}}
      \sv{g} \cOpApp (\csem{f}_{\ets} \cOpApp \addCost{\appCost(\vmapPair \cOpApp \cOpPair{\sv{g}}{\sv{g}}, \cOpPair{\sv{x_1}}{\sv{x_2}})}{\cOpPair{\sv{x_1}}{\sv{x_2}}})\\
      = \addCost{c}{(\csem{f}_{\ets} \cOpApp (\vmapPair \cOpApp \cOpPair{\sv{g}}{\sv{g}} \cOpApp \cOpPair{\sv{x_1}}{\sv{x_2}}))}
    \end{array}
  \end{array}
  \]
  and thus:
  \[\forall \tau_1, \tau_2\text{ types}, g\cta \tau_1 \to \tau_2, t\cta(\tau_1,\tau_1).~g~(f~t)\sqsubseteq f~(\mapPair~(g, g)~t)\]
\end{example}

In order to also be able to deal with example types involving lists, we define $\vmapList=\csem{\mapList}_{\ets}$ for $\mapList$ as given in Section~\ref{sec:calculus}.
Then, we give analogues of Propositions~\ref{prop:relspec}/\ref{prop:relspec:pair} and Corollaries~\ref{cor:relspec}/\ref{cor:relspec:pair}, given sets $S_1$ and $S_2$, a $\Cost{\cdot}$-lifted function $\sv{g} \in \Cost{\fspace{S_1}{\Cost{S_2}}}$, and $\Cost{\cdot}$-lifted values $\sv{x_1}, \ldots, \sv{x_n} \in \Cost{S_1}$, with $n\in \N$.
\begin{proposition}\label{prop:relspec:list}
  We have 
  $(\sv{xs}, \sv{ys}) \in \cllift{\Cost{\rR^{\sv{g}}_{\sv{x_1}, \ldots, \sv{x_n}}}}$ if and only if there exist $m\in\N$, $i_1, \ldots, i_m \in \{1,\ldots, n\}$, and $\svc{c} \in \Z$ such that $\sv{xs} = \addCost{\svc{c}}{\addCost{\appCost(\vmapList \cOpApp \sv{g}, \clist{\sv{x_{i_1}}, \ldots, \sv{x_{i_m}}})}{\clist{\sv{x_{i_1}}, \ldots, \sv{x_{i_m}}}}}$ and $\sv{ys} = \addCost{\svc{c}}{(\vmapList \cOpApp \sv{g} \cOpApp \clist{\sv{x_{i_1}}, \ldots, \sv{x_{i_m}}})}$.
\end{proposition}
\begin{corollary}\label{cor:relspec:list}
  If $(\sv{xs}, \sv{ys}) \in \cllift{\Cost{\rR^{\sv{g}}_{\sv{x_1}, \ldots, \sv{x_n}}}}$, then there exist $m\in\N$ and $i_1, \ldots, i_m \in \{1,\ldots, n\}$ such that $\vmapList \cOpApp \sv{g} \cOpApp \sv{xs} = \addCost{\appCost(\vmapList \cOpApp \sv{g}, \clist{\sv{x_{i_1}}, \ldots, \sv{x_{i_m}}})}{\sv{ys}}$ and $\getVal{\clist{\sv{x_{i_1}}, \ldots, \sv{x_{i_m}}}}=\getVal{\sv{xs}}$.
\end{corollary}
Note that the final conclusion in the corollary, $\getVal{\clist{\sv{x_{i_1}}, \ldots, \sv{x_{i_m}}}}=\getVal{\sv{xs}}$, keeps a bit more information than we have cared to keep in Corollaries~\ref{cor:relspec} and~\ref{cor:relspec:pair}.
The reason is that this information will be useful in Example~\ref{ex:reverse} below.
\begin{example}
  Let some term $f$ be given with $\alpha \vdash f \cta [\alpha] \to \Int$.
  Using Corollary~\ref{cor:para}, the definition of the logical relation, and Proposition~\ref{prop:relspec:list} for $\rR^{\sv{g}}_{\sv{x_1}, \ldots, \sv{x_n}}$, we get:
  \[
  \begin{array}{@{}l@{}}
    \forall S_1, S_2\text{ sets}, \sv{g} \in \Cost{\fspace{S_1}{\Cost{S_2}}}, n\in\N,\sv{x_1},\ldots,\sv{x_n} \in \Cost{S_1}.\\
    ~
    \csem{f}_{\ets} \cOpApp \addCost{\appCost(\vmapList \cOpApp \sv{g}, \clist{\sv{x_1},\ldots,\sv{x_n}})}{\clist{\sv{x_1},\ldots,\sv{x_n}}} = \csem{f}_{\ets} \cOpApp (\vmapList \cOpApp \sv{g} \cOpApp \clist{\sv{x_1},\ldots,\sv{x_n}})
  \end{array}
  \]
  and thus:
  \[\forall \tau_1, \tau_2\text{ types}, g\cta \tau_1 \to \tau_2, t\cta[\tau_1].~f~t\sqsubseteq f~(\mapList~g~t)\]
\end{example}
\begin{example}\label{ex:reverse}
  Let some term $f$ be given with $\alpha \vdash f \cta [\alpha] \to [\alpha]$.
  Using Corollary~\ref{cor:para}, the definition of the logical relation, and Proposition~\ref{prop:relspec:list} and Corollary~\ref{cor:relspec:list} for $\rR^{\sv{g}}_{\sv{x_1}, \ldots, \sv{x_n}}$, plus simplification, we get:
  \[
  \begin{array}{@{}l@{}}
    \forall S_1, S_2\text{ sets}, \sv{g} \in \Cost{\fspace{S_1}{\Cost{S_2}}}, n\in\N,\sv{x_1},\ldots,\sv{x_n} \in \Cost{S_1}.~\exists m \in \N, i_1, \ldots, i_m \in \{1, \ldots, n\}.\\
    ~
    \begin{array}[t]{@{}l@{}}
      \addCost{\appCost(\vmapList \cOpApp \sv{g}, \clist{\sv{x_1},\ldots,\sv{x_n}})}{(\vmapList \cOpApp \sv{g} \cOpApp (\csem{f}_{\ets} \cOpApp \clist{\sv{x_1},\ldots,\sv{x_n}}))}\\
      = \addCost{\appCost(\vmapList \cOpApp \sv{g}, \clist{\sv{x_{i_1}}, \ldots, \sv{x_{i_m}}})}{(\csem{f}_{\ets} \cOpApp (\vmapList \cOpApp \sv{g} \cOpApp \clist{\sv{x_1},\ldots,\sv{x_n}}))}\\
      \mathop{\wedge} \getVal{\clist{\sv{x_{i_1}}, \ldots, \sv{x_{i_m}}}}=\getVal{\csem{f}_{\ets} \cOpApp \clist{\sv{x_1}, \ldots, \sv{x_n}}}
    \end{array}
  \end{array}
  \]
  In order to continue now and derive a statement about the relative efficiencies of $\mapList~g~(f~t)$ and $f~(\mapList~g~t)$, for types $\tau_1, \tau_2$, function $g\cta \tau_1 \to \tau_2$, and list $t\cta[\tau_1]$, we would need further information about $\appCost(\vmapList \cOpApp \sv{g}, \clist{\sv{x_1},\ldots,\sv{x_n}})$ and $\appCost(\vmapList \cOpApp \sv{g}, \clist{\sv{x_{i_1}}, \ldots, \sv{x_{i_m}}})$.
  This cannot be provided generally, but a number of useful observations is possible.
  For example, we know that the elements $\sv{x_{i_1}}, \ldots, \sv{x_{i_m}}$ form a subset of $\{\sv{x_1},\ldots,\sv{x_n}\}$, and hence that evaluation of $\mapList~g~(f~t)$ does not incur $g$-costs on elements other than those already encountered during evaluation of $f~(\mapList~g~t)$, though of course a different selection and multiplicities are possible.
  Moreover, if we assume that $g$ (actually, $\sv{g}$) is equally costly on every element of $t$ (on every $\sv{x_i}$), or indeed on every term of type $\tau_1$ (on every element of $\Cost{S_1}$), then we can reduce the question about the relative efficiency of $\mapList~g~(f~t)$ and $f~(\mapList~g~t)$ to one about the relative length of $t$ and $f~t$, to which an answer might be known statically by some separate analysis.
  Also, note that with some extra effort it would even have been possible to explicitly get our hands at the existentially quantified $m$ and $i_1, \ldots, i_m$, namely to establish that $[\svv{i_1}, \ldots, \svv{i_m}]=\getVal{\csem{f}_{\ets} \cOpApp ([\svv{1},\ldots,\svv{n}],0)}$.
\end{example}

Let us also briefly comment on applying our machinery to an automatic program transformation that is used in a production compiler~\citep[though in a call-by-need setting, the mainstream Glasgow Haskell Compiler]{glp93}.
The cost-insensitive content of the underlying ``short-cut fusion'' rule, typically proved via a standard free theorem, can be expressed in our setting as follows, for every choice of types $\tau$ and $\tau'$, polymorphic function $g \cta (\tau \to \alpha \to \alpha) \to \alpha \to \alpha$, and $k\cta \tau\to\tau'\to\tau'$ and $z::\tau'$:
\[\getVal{\csem{\lFold{k,z,g[[\tau]/\alpha]~(\lambda x \cta \tau. \lambda \var{xs} \cta [\tau]. x \cons \var{xs})~\nil_\tau}}_{\ets}} = \getVal{\csem{g[\tau'/\alpha]~k~z}_{\ets}}\]
The desirable statement, and certainly the intuitive assumption by which application of short-cut fusion in a compiler is usually justified, would be:
\begin{equation}
  \label{eq:shortcut}
  \csem{\lFold{k,z,g[[\tau]/\alpha]~(\lambda x \cta \tau. \lambda \var{xs} \cta [\tau]. x \cons \var{xs})~\nil_\tau}}_{\ets} \sqsupseteq \csem{g[\tau'/\alpha]~k~z}_{\ets}
\end{equation}
We could even hope to quantify the $c\geq 0$ such that $\csem{\lFold{k,z,\ldots
}}_{\ets}=\addCost{c}{\csem{g[\tau'/\alpha]~k~z}_{\ets}}$ holds, possibly expressing $c$ in terms of the length of the intermediate list $\getVal{\csem{g[[\tau]/\alpha]~(\lambda x \cta \tau. \lambda \var{xs} \cta [\tau]. x \cons \var{xs})~\nil_\tau}_{\ets}}$.
But, maybe surprisingly, (\ref{eq:shortcut}) does \emph{not} actually hold in general.
The reason is that $g$ may ``use'' its arguments for other things than for creating its output.
For example, with $\tau=\Int$, $g$ could be the function $\lambda k\cta \Int \to \alpha \to \alpha.\lambda z\cta \alpha.(\lambda x\cta\alpha.z)~(k~5~z)$.
Then:
\begin{enumerate}
\item\label{item:1}
  On the one hand, $\lFold{k,z,\ldots}$ incurs no costs at all from applying a concrete $k\cta \Int\to\tau'\to\tau'$ to any values, because $g[[\Int]/\alpha]$ is only applied to $(\lambda x \cta \Int. \lambda \var{xs} \cta [\Int]. x \cons \var{xs})$ and $\nil_\Int$ during its evaluation, leading to the empty list as intermediate result which is then processed by the~$\plFold$.

\item
  On the other hand, $g[\tau'/\alpha]~k~z$ does incur costs for evaluating the application $k~5~z$, even though the resulting value is eventually discarded in $(\lambda x\cta\alpha.z)~(k~5~z)$.
  Moreover, since we are free to choose $k$ (and~$z$) however we want, we are certainly free to make that application $k~5~z$ arbitrarily more costly than the corresponding application $(\lambda x \cta \Int. \lambda \var{xs} \cta [\Int]. x \cons \var{xs})~5~\nil_\Int$ contributing to the cost of~\ref{item:1}.~above.
\end{enumerate}
Hence, the right-hand side of~(\ref{eq:shortcut}) can be made arbitrarily more costly than its left-hand side.
(The same behavior can be provoked in Haskell using the $\var{seq}$-primitive.)
It is possible to constrain $g$ in such a way that~(\ref{eq:shortcut}) actually holds, and indeed all ``reasonable'' functions to be used in short-cut fusion can be expected to satisfy the condition thus imposed on $g$, but spelling out the details is left for future work.

\section{Conclusion}

We have developed a notion of relational parametricity that incorporates information about call-by-value evaluation costs, and thus allows to derive quantitative statements about runtime from function types.
The mechanics of deriving statements that way are a bit more involved than in the purely extensional setting, but we are optimistic that automation like for \url{http://www-ps.iai.uni-bonn.de/cgi-bin/free-theorems-webui.cgi} \citep{boe07} is possible here as well.

As already mentioned, the exact way in which we assign costs to different program constructs does not appear to impact the overall approach much.
Hence, we could also work with more detailed and realistic measures, as for example in the work of~\citet{lg01}.
Of course, we are also interested in moving from a call-by-value setting to a call-by-name/need one, and in extending the results for our calculus to a calculus with general recursion.

\renewcommand{\bibsection}{\section{References}}



\appendix

\section{Proof Sketch of Theorem~\ref{thm:para}}\label{app:para}

  The proof is by induction over the typing derivation, \ie we have to consider the derivation rules in Figure~\ref{fig:typingrules}.
  In the proof we use the same names for the environments as in Theorem~\ref{thm:para} (\ie $\rho$, $\sigma_1$, $\sigma_2$) and assume the conditions on them that are given in Theorem~\ref{thm:para} are satisfied.
  We show just three cases.

  In the case 
  \[\Gamma,x\cta\tau\vdash x\cta\tau
  \]
  the second condition in Theorem~\ref{thm:para} ensures that $(\sigma_1(x), \sigma_2(x)) \in \logrel{\tau}{\rho}$ and hence 
  it holds that $(\csem{x}_{\tsone}, \csem{x}_{\tstwo})\pbr=((\sigma_1(x), \svc{0}) , (\sigma_2(x), \svc{0}))$ is in $\Cost{\logrel{\tau}{\rho}}$.

  In the case
  \[
  \AXC{$\Gamma,x\cta\tau_1\vdash t\cta\tau_2$}
  \UIC{$\Gamma\vdash(\lambda x\cta\tau_1.t)\cta\tau_1\to\tau_2$}
  \DP
  \]
  we have
  \begin{align*}
          & (\csem{\lambda x \cta \tau_1. t}_{\tsone}, \csem{\lambda x \cta \tau_1. t}_{\tstwo}) \in \Cost{\logrel{\tau_1 \to \tau_2}{\rho}} \\
    \eqa  & ((\uplambda \svv{v}. \addCost{\svc{1}}{\csem{t}_{\tsone[x \mapsto \svv{v}]}}, \svc{0}), (\uplambda \svv{v'}. \addCost{\svc{1}}{\csem{t}_{\tstwo[x \mapsto \svv{v'}]}}, \svc{0})) \in \Cost{\logrel{\tau_1 \to \tau_2}{\rho}} \\
    \eqa  & (\uplambda \svv{v}. \addCost{\svc{1}}{\csem{t}_{\tsone[x \mapsto \svv{v}]}}, \uplambda \svv{v}. \addCost{\svc{1}}{\csem{t}_{\tstwo[x \mapsto \svv{v}]}}) \in \logrel{\tau_1 \to \tau_2}{\rho} \\
    \eqa  & \forall (\svv{v}, \svv{v'}) \in \logrel{\tau_1}{\rho}.~ (\addCost{\svc{1}}{\csem{t}_{\tsone[x \mapsto \svv{v}]}}, \addCost{\svc{1}}{\csem{t}_{\tstwo[x \mapsto \svv{v'}]}}) \in \Cost{\logrel{\tau_2}{\rho}} \\
    \eqa  & \forall (\svv{v}, \svv{v'}) \in \logrel{\tau_1}{\rho}.~ (\csem{t}_{\tsone[x \mapsto \svv{v}]}, \csem{t}_{\tstwo[x \mapsto \svv{v'}]}) \in \Cost{\logrel{\tau_2}{\rho}}
  \end{align*}
  where the last line is the induction hypothesis.

  In the case
  \[
  \AXC{$\Gamma\vdash t_1\cta\tau_1\to\tau_2$}
  \AXC{$\Gamma\vdash t_2\cta\tau_1$}
  \BIC{$\Gamma\vdash (t_1~t_2)\cta\tau_2$}
  \DP
  \]
  we reason as follows:
  \begin{align*}
    &  (\csem{t_1~t_2}_{\tsone}, \csem{t_1~t_2}_{\tstwo}) \in \Cost{\logrel{\tau_2}{\rho}} \\
    \eqa & (\csem{t_1}_{\tsone} \cOpApp \csem{t_2}_{\tsone}, \csem{t_1}_{\tstwo} \cOpApp \csem{t_2}_{\tstwo}) \in \Cost{\logrel{\tau_2}{\rho}} \\
    \impl & \forall (\sv{x}, \sv{y}) \in \Cost{\logrel{\tau_1}{\rho}}.~ (\csem{t_1}_{\tsone} \cOpApp \sv{x}, \csem{t_1}_{\tstwo} \cOpApp \sv{y}) \in \Cost{\logrel{\tau_2}{\rho}} \\
    \impl & (\csem{t_1}_{\tsone}, \csem{t_1}_{\tstwo}) \in \Cost{\logrel{\tau_1 \to \tau_2}{\rho}}
  \end{align*}
  The last line is the first induction hypothesis, the last implication is by Lemma~\ref{lem:lrfun}, and the second last implication by the second induction hypothesis. 

\end{document}